\documentclass[runningheads,a4paper]{llncs}
\pdfoutput=1
\usepackage{amsmath,cite}
\usepackage{mathrsfs}
\usepackage{amssymb,color}
\usepackage{graphicx,subfigure}
\usepackage{url}
\newcommand{\keywords}[1]{\par\addvspace\baselineskip
\noindent\keywordname\enspace\ignorespaces#1}
\usepackage[generate,shellescape,ps2eps]{abc}

\begin{document}

\mainmatter  

\title{Music transcription modelling and composition using deep learning}

\titlerunning{Music transcription modelling and composition using deep learning}

%
%

\author{Bob L. Sturm\inst{1}, Jo\~ao Felipe Santos\inst{2}, Oded Ben-Tal\inst{3} and 
Iryna Korshunova\inst{4}\thanks{The authors would like to thank Dr. Nick Collins, 
Jeremy Keith creator and host of {\tt thesession.org}, and its many contributors.}} 
%
\authorrunning{Sturm, Santos, Ben-Tal and Korshunova}

\institute{Centre for Digital Music, Queen Mary University of London \and INRS-EMT, Montreal Canada \and Music Department, Kingston University, UK\and ELIS, Ghent University, Belgium}

%
%

\maketitle

\begin{abstract}
We apply deep learning methods, specifically 
long short-term memory (LSTM) networks,
to music transcription modelling and composition.
We build and train LSTM networks
using approximately 23,000 music transcriptions expressed 
with a high-level vocabulary (ABC notation),
and use them to generate new transcriptions.
Our practical aim is to create music transcription models 
useful in particular contexts of music composition.
We present results from three perspectives:
1) at the population level, comparing descriptive statistics of 
the set of training transcriptions and generated transcriptions;
2) at the individual level,
examining how a generated transcription reflects
the conventions of a music practice in the training transcriptions (Celtic folk);
3) at the application level,
using the system for idea generation in music composition.
We make our datasets, software and sound examples
open and available: \url{https://github.com/IraKorshunova/folk-rnn}.


\end{abstract}

\section{Introduction}
The application of artificial neural networks to music modelling,
composition and sound synthesis is not new, e.g., 
\cite{Dolson1989a,Todd1991a,Tudor1992a,Leman1992,Griffith1999};
but what is new is the unprecedented accessibility to resources:
from computational power to data, 
from superior training methods to 
open and reproducible research.
This accessibility is a major reason ``deep learning'' methods \cite{Deng2014,LeCun2015} 
are advancing far beyond state of the art results in many applications of machine learning,
for example, image content analysis \cite{Krizhevsky2012},
speech processing \cite{Hinton2012} and recognition \cite{Graves2013a}, 
text translation \cite{Sutskever2014},
and, more creatively, artistic style transfer \cite{Gatys2015},
and Google's Deep Dream.\footnote{\url{https://en.wikipedia.org/wiki/DeepDream}}
As long as an application domain is ``data rich,''
deep learning methods stand to make substantial contributions.

Deep learning is now being applied to music data,
from analysing and modelling the content of sound recordings \cite{Lee2009,Spiliopoulou2011,Yang2011b,Humphrey2013,Zhang2014,Sigtia2014,Sturm2014,Kereliuk2015a},
to generating new music \cite{Coca2011,Spiliopoulou2011,Boulanger-Lewandowski2014}.
Avenues for exploring these directions are 
open to many since powerful software tools
are free and accessible, e.g., Theano \cite{Bergstra2010},
and compatible computer hardware, e.g., graphical processing units, is inexpensive.
This has led to a variety of ``garden shed experiments''
described in a timely manner on various public web logs.\footnote{\scriptsize\url{deeplearning.net/tutorial/rnnrbm.html} \\ \url{www.hexahedria.com/2015/08/03/composing-music-with-recurrent-neural-networks} \\ \url{www.wise.io/tech/asking-rnn-and-ltsm-what-would-mozart-write} \\ 
\url{elnn.snucse.org/sandbox/music-rnn}}
The work we describe here moves beyond our informal experiments\footnote{\scriptsize\url{
highnoongmt.wordpress.com/2015/05/22/lisls-stis-recurrent-neural-networks-for-folk-music-generation}}
to make several contributions. 

In particular, we build long short-term memory (LSTM) networks
having three hidden layers of 512 LSTM blocks each,
and train them using approximately 23,000 music transcriptions  
expressed with a textual vocabulary (ABC notation).
We use this data because it is available,
high-level with regards to the music it transcribes,
and quite homogeneous with regards to 
the stylistic conventions of the music
(it is crowd-sourced
by musicians that play ``session'' music, e.g., 
Celtic, Morris, etc.).
We take two approaches to training our models:
one is character based, in which the system 
builds a model of joint probabilities of 
each textual character given the previous 50 characters;
the other is ``token'' based, in which the system computes
the joint probability of each token (which can be more than one character) 
given all previous tokens of a transcription.
The result of training is a generative system
that outputs transcriptions resembling those in the training material.
Our practical aim is to create music transcription models 
that are useful in particular contexts of music composition,
within and outside stylistic conventions particular to the training data.

In the next section, we review deep learning and LSTM,
as well as past work applying such networks to music modelling and generation.
Section 3 describes the specific models we build.
In section 4, we analyse our generative models from three perspectives:
1) we compare the descriptive statistics of 
the set of training transcriptions and the generated transcriptions of a model;
2) we examine how a generated transcription reflects
the conventions of a music practice in the training transcriptions (e.g., Celtic folk \cite{Hillhouse2005a});
3) we use a model for music composition
outside the stylistic conventions of the training data.
Our contributions include extending similar past work
by using much larger networks and much more data (see Sec. 2.2),
by studying the actual application of our models for assisting in music composition,
and by making our datasets and software freely available.



%
\vspace{-0.1in}
\section{Background}

\subsection{Long short term memory (LSTM) networks}
A deep neural network is one that has more than one hidden layer
of units (neurons) between its input and output layers \cite{LeCun2015}.
Essentially, a neural network transforms an input
by a series of cascaded non-linear operations.
A recurrent neural network (RNN) is any neural network possessing a directed
connection from the output of at least one unit into the input of another unit
located at a shallower layer than itself (closer to the input). 
A deep RNN is a stack of several RNN layers, where each hidden layer 
generates an output sequence 
that is then used as a sequential input for the deeper layer. 
With deeper architectures, one
expects each layer of the network to be able to learn higher level
representations of the input data and its short- and long-term relationships.

The recurrence (feedback) present in an RNN 
allows it to take into account its past inputs
together with new inputs.
Essentially, an RNN predicts a sequence of symbols
given an input sequence. 
Training it entails modifying the parameters of its transformations
to diminish its prediction error for a dataset of known sequences.
The basic recurrent structure, however, 
presents problems related to exploding and
vanishing gradients during the training procedure \cite{Hochreiter1998,Pascanu2013a}, 
which can result in a lack of convergence of solutions. 
These problems can be circumvented by defining 
the hidden layer activation function 
in a smart way. 
One such approach defines long short term memory (LSTM) ``cells'',
which increases the number of parameters to be estimated in training, 
but controls the flow of information in and out of each cell 
to greatly help with convergence \cite{Hochreiter1997,Graves2013a}. 

Though RNN and LSTM are not new,
recent advances in efficient training algorithms and the prevalence of data
have led to great success when they are applied to sequential data processing
in many domains, e.g., continuous handwriting \cite{Graves2013b}, speech
recognition \cite{Graves2013a}, and machine translation \cite{Sutskever2014}.
In the next subsection, we describe past applications of 
recurrent networks to music transcription modelling and generation.

\vspace{-0.1in}
\subsection{Music modelling and generation using RNN and LSTM}
Describing music as a sequence of symbols 
makes RNN immediately applicable to model it \cite{Todd1989,Mozer1994a,Griffith1999,Chen2001,Eck2002a,Franklin2006a,Eck2008a,Boulanger-Lewandowski2014}.
The RNN built and tested by Todd \cite{Todd1989} 
consist of an input layer having 19 units,
a single hidden layer with 8-15 units,
and an output layer of 15 units.
One unit in each of the input and output layers
is the ``note begin'' state;
14 other units represent pitch,
one each from D4 to C6 (no accidentals).
Four other input units identify 
a specific monophonic training melody,
of which there are four,  
each 34 notes long.
Todd divides time such that
each time-step of the model
represents an eighth-note duration.

Mozer \cite{Mozer1994a} builds RNN
to model and generate melody
using a distributed approach to music encoding.
These systems generate output at the note level
rather than at uniform time steps.
Each pitch is encoded based on its fundamental frequency, 
chromatic class, and position in the circle of fifths.
Note duration is encoded using a similar approach.
Chordal accompaniment is encoded based on the pitches present.
Some input units denote time signature, key, and downbeats.
Mozer's RNN employs a single hidden layer with $\mathcal{O}(10)$ units.
Training material include artificial sequences (scales, random walks),
10 melodies of J. S. Bach (up to 190 notes long),
25 European folk melodies,
and 25 waltzes. 
Mozer finds 
these systems can succeed when it comes to
modelling local characteristics of melody, e.g., stepwise motions,
but fail to capture longer structures, e.g., phrasing, rhythm, resolution.

The finding of Mozer provided motivation 
for the work of Eck and Schmidhuber \cite{Eck2002a},
the first to apply LSTM networks to music modelling and generation.
Similar to Todd \cite{Todd1989},
they employ 
a local music encoding approach with 13 units representing 13 pitches (chromatic octave),
and divide time using a minimum duration, e.g., sixteenth note.
They also use 12 input units to designate pitches in an accompanying harmony.
The hidden layer consists of two blocks of 8 LSTM cells each,
with one block devoted to melody and the other to harmony.
They make recurrent connections from the melody block 
to the harmony block, but not the other way around.
They train the system on 6 minutes of 
12-bar blues melodies with chord accompaniment, 
encoded at 8 time steps per bar.
Each training song is 96 time steps long.
Compared with the results of Mozer \cite{Mozer1994a},
Eck and Schmidhuber find that the LSTM network demonstrates an ability to 
model and reproduce long term conventions of this style.
In a similar direction, Franklin \cite{Franklin2006a} 
models jazz melodies and harmonic accompaniment using LSTM networks,
but using a distributed music encoding similar to that used by Mozer \cite{Mozer1994a}.

Chen and Miikkulainen \cite{Chen2001} ``evolve'' an RNN 
using fitness functions that quantify the success of 
a melody along different qualities, e.g., short-term movement,
and pitch and rhythm diversity.
They define some of these constraints to favor the melodic style of Bartok, e.g., pentatonic modes.
Chen and Miikkulainen appear to encode a melody measure wise, 
using 16 pairs of pitch interval and duration.
Output units are read in a linear fashion,
with pairs of interval and duration,
until the length of a full measure is completed.

Eck and Lapamle \cite{Eck2008a} applied LSTM networks
to modelling long-term conventions of transcriptions of Irish folk music.
Their music encoding divides time into eighth-note durations,
with each note (between C3-C5) and chord getting its own bit.
A novel aspect is that the LSTM network input is a linear combination 
of the current note and past notes from metrically related times, 
e.g., 4, 8, and 12 measures before.
They train their systems on transcriptions of reels
transposed to the same key:
56 from {\tt http://thesession.org} (the source of our training data), 
and 435 from another database.
They take care to reset the training error propagation at transcription boundaries.


More recently, Boulanger-Lewandowski et al. \cite{Boulanger-Lewandowski2012a}
apply RNN to modelling and generating polyphonic music transcriptions.
They encode music by absolute pitch (88 notes from A0 to C8),
quantised to the nearest quarter note duration.
They train several networks on different datasets,
e.g., Classical piano music, folk tunes, Bach chorals,
and find the generated music lacks long-term structure.
(We hear such results in the music produced 
in the links of footnote 4 above.)


\vspace{-0.15in}
\section{Creating our generative LSTM networks}\vspace{-0.05in}
All our LSTM networks have the same architecture, 
but operate over different vocabularies and are trained differently.
One kind we build, which we term {\em char-rnn},
operates over a vocabulary of single characters,
and is trained on a continuous text file.
The second kind we build, {\em folk-rnn},
operates over a vocabulary of transcription tokens,
and is trained on single complete transcriptions.
We next discuss our training data,
and then the architecture and training of our systems,
and finally how we use them to generate new transcriptions.

%
\subsection{Music transcription data}\label{sec:trainingdata}
Our transcription data comes from 
a weekly repository
of { \url{https://thesession.org/}},\footnote{\url{https://github.com/adactio/TheSession-data}} 
an on-line platform
for sharing and discussing music 
played in ``traditional music sessions'' 
(often Celtic and Morris).
The collection does not include just music transcriptions, 
but also discussions, jokes, accompaniment suggestions, and so on.
All transcriptions are expressed in  
``ABC'' notation.\footnote{\url{http://abcnotation.com/wiki/abc:standard:v2.1}}
Entries in the repository look like the following real examples:

{\small
\begin{verbatim}
3038,3038,"A Cup Of Tea","reel","4/4","Amixolydian","|:eA (3AAA g2 fg | 
eA (3AAA BGGf|eA (3AAA g2 fg|1afge d2 gf:|2afge d2 cd||
|:eaag efgf|eaag edBd|eaag efge|afge dgfg:|","2003-08-28 21:31:44","dafydd"
3038,21045,"A Cup Of Tea","reel","4/4","Adorian","eAAa ~g2fg|eA~A2 BGBd| 
eA~A2 ~g2fg|1af (3gfe dG~G2:|2af (3gfe d2^cd||eaag efgf|
eaag ed (3Bcd|eaag efgb|af (3gfe d2^cd:|","2013-02-24 13:45:39",
"sebastian the megafrog"
\end{verbatim}}

\normalsize
\noindent An entry begins with two identifiers, followed by the title,
tune type, meter, key, ABC code, date, and contributing user.
Contributions vary in detail, with some being quite elaborate,
e.g., specifying ornamentation, grace notes, slurs and chords.
Most transcriptions are monophonic, but some do specify multiple voices. 
Many transcriptions have improper ABC formatting, are missing bar lines, 
have redundant accidentals, miscounted measures, and so on.

We create data for training our {\em char-rnn} model in the following way.
We keep only five ABC fields (title, meter, key, unit note length, and transcription),
and separate each contribution by a blank line. 
The two entries above thus become:

{\small
\begin{verbatim}
T: A Cup Of Tea
M: 4/4
L: 1/8
K: Amix
|:eA (3AAA g2 fg|eA (3AAA BGGf|eA (3AAA g2 fg|1afge d2 gf:|2afge d2 cd||
|:eaag efgf|eaag edBd|eaag efge|afge dgfg:|

T: A Cup Of Tea
M: 4/4
L: 1/8
K: Ador
eAAa ~g2fg|eA~A2 BGBd|eA~A2 ~g2fg|1af (3gfe dG~G2:|2af (3gfe d2^cd||
eaag efgf|eaag ed (3Bcd|eaag efgb|af (3gfe d2^cd:|
\end{verbatim}}

\normalsize
\noindent This leaves us with a text file having 
13,515,723 characters in total,
and 47,924 occurrences of {\tt T:}.\footnote{This is not the number of transcriptions in the data
because it also includes such things as user discussions and accompaniment suggestions
for particular tunes.}
There are 135 unique characters, e.g., ``A'', ``:'', and ``{\tt \textasciicircum}'',
each of which becomes an element of the vocabulary
for our {\em char-rnn} model.

We create data for training our {\em folk-rnn} model in the following way.
We remove title fields and ornaments.
We remove all transcriptions that have fewer than 7 measures
when considering repetitions (to remove
contributions that are not complete transcriptions, 
but transcriptions of suggested endings, variations, etc.).
We remove all transcriptions that have more than one meter or key.\footnote{By converting the remaining transcriptions to MIDI,
we find the following:
78,338 measures of incorrect lengths (miscounting of notes, 
among 725,000+ measure symbols),
4,761 unpaired repeat signs,
and 3,057 incorrect variant endings (misspecified repetitions).
We do not attempt to correct these problems.} 
We transpose all remaining transcriptions (23,636) to a key with root C.
All transcriptions are thus in one of the four modes
(with percentage shown in parens):
major (67\%), minor (13\%), dorian (12\%), and mixolydian (8\%).
We impose a transcription token vocabulary ---
each token consists of one or more characters ---
for the following seven types (with examples in parens):
meter (``{\tt M:3/4}''), key (``{\tt K:Cmaj}''), measure (``{\tt :|}'' and ``{\tt |1}''),
pitch (``{\tt C}'' and ``{\tt \textasciicircum c'}''),
grouping (``{\tt (3}''), duration (``{\tt 2}'' and ``{\tt /2}''),
and transcription (``{\tt <s>}'' and ``{\tt <$\backslash$s>}'').
The two transcriptions above are thus expressed as

\vspace{-0.05in}
{\footnotesize
\begin{verbatim}
<s> M:4/4 K:Cmix |: g c (3 c c c b 2 a b | g c (3 c c c d B B a | g c (3 
c c c b 2 a b |1 c' a b g f 2 b a :| |2 c' a b g f 2 e f |: g c' c' b g 
a b a | g c' c' b g f d f | g c' c' b g a b g | c' a b g f b a b :| <\s>
<s> M:4/4 K:Cdor g c c c' b 2 a b | g c c 2 d B d f | g c c 2 b 2 a b |1 
c' a (3 b a g f B B 2 :| |2 c' a (3 b a g f 2 =e f | g c' c' b g a b a | g 
c' c' b g f (3 d e f | g c' c' b g a b d' | c' a (3 b a g f 2 =e f :| <\s>
\end{verbatim}}\vspace{-0.05in}

\normalsize
\noindent Our dataset has 4,056,459 tokens,
of which 2,816,498 are pitch, 602,673 are duration, and 520,290 are measure.
A majority of the 23,636 transcriptions consists of 150 tokens or fewer;
and 75\% have no more than 190.
There are 137 unique tokens, 
each of which becomes a vocabulary element 
for our {\em folk-rnn} model.

\vspace{-0.1in}
\subsection{Architecture}
Each LSTM network we build has three hidden layers with 512 LSTM blocks each,
and a number of input and output units equal to the number of 
characters or tokens in its vocabulary.
We encode our transcriptions in a local fashion, 
like in \cite{Todd1989, Eck2002a},
where each element in the vocabulary 
is mapped to an input and output unit.
(This is also called ``one-hot encoding''.)
The output of each network is a probability distribution over its vocabulary.
The total number of parameters in our {\em char-rnn} model is 5,585,920;
and that in our {\em folk-rnn} model is 5,621,722.
%



%
\vspace{-0.1in}
\subsection{Training}
We build and train our {\em char-rnn} model
using the ``char-rnn'' implementation.\footnote{\small\url{https://github.com/karpathy/char-rnn}}
This employs the RMSprop algorithm\footnote{T. Tieleman and G. Hinton,
``Divide the gradient by a running average of its recent magnitude,''
lecture 6.5 of Coursera ``Neural Networks for Machine Learning,'' 2012.} 
using minibatches of 50
samples containing 50 characters each,
and a gradient clipping strategy to avoid the exploding gradients problem in the LSTMs.
We initialise the learning rate to 0.002,
and apply a decay rate of 0.95 after the first 10 epochs.
We build and train our \emph{folk-rnn} model 
using our own implementation. 
This also employs the RMSprop algorithm,
but with minibatches of 64 parsed transcriptions each. 
Since transcriptions in the dataset have
different lengths (in number of tokens), we generate minibatches using a
bucketing strategy, which places together in a minibatch sequences with
approximately the same length, pads them to the maximum length using a ``null''
token, and then use a masking strategy to ignore null tokens when computing
outputs and the loss function. 
We begin training with a learning rate of 0.003,
and a rate decay of 0.97 applied after the 20 first epochs.  

For both models, we clip gradients outside $[-5,5]$ to the limits,
and employ a dropout rate of 0.5 after each LSTM hidden layer. 
We train each model for 100 epochs in total.
We use 95\% of the dataset as training data and 5\% as validation data
(the latter for measuring progress in predicting characters or tokens).
Through training, our {\em char-rnn} model 
learns a ``language model'' to produce ABC characters.
On the contrary, our {\em folk-rnn} model
learns a language model in a vocabulary more specific to transcription,
i.e., a valid transcription begins with {\tt <s>}, then a time signature token, a key token, 
and then a sequence of tokens from 4 types.
Our {\em folk-rnn} model does not embody
the ambiguity of meaning that {\em char-rnn} does,
e.g., that {\tt C} can mean a pitch, part of a pitch ({\tt \textasciicircum C}),
a letter in a title ({\tt A Cup of Tea}),
or part of a key designation ({\tt K:Cmin}).

\vspace{-0.1in}
\subsection{Generating transcriptions}
With our trained models, it is a simple matter to have them generate output:
we just sample from the probability distribution output by the model
over its vocabulary, and use each selected vocabulary element as subsequent input.
We can initialise the internal state of each model either randomly,
or by inputing a valid ``seed'' sequence (e.g., beginning with {\tt <s>}).
Repeating the sampling process for $N$ timesteps
produces $N$ characters/tokens in addition to the seed sequence.

\vspace{-0.1in}
\section{Demonstrations of our generative LSTM networks}

\begin{figure}[t]
  \centering
\subfigure{\includegraphics[width=0.85\textwidth]{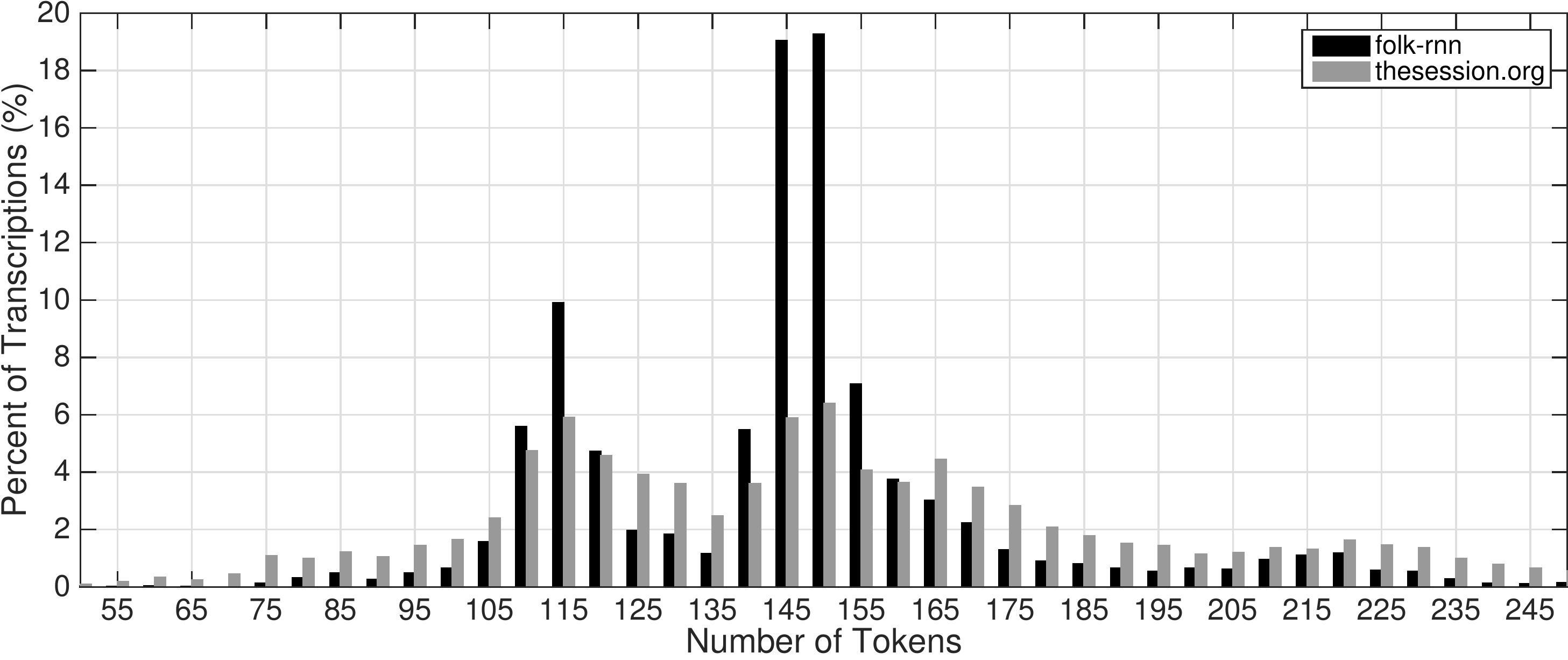}}
\subfigure{\includegraphics[width=0.85\textwidth]{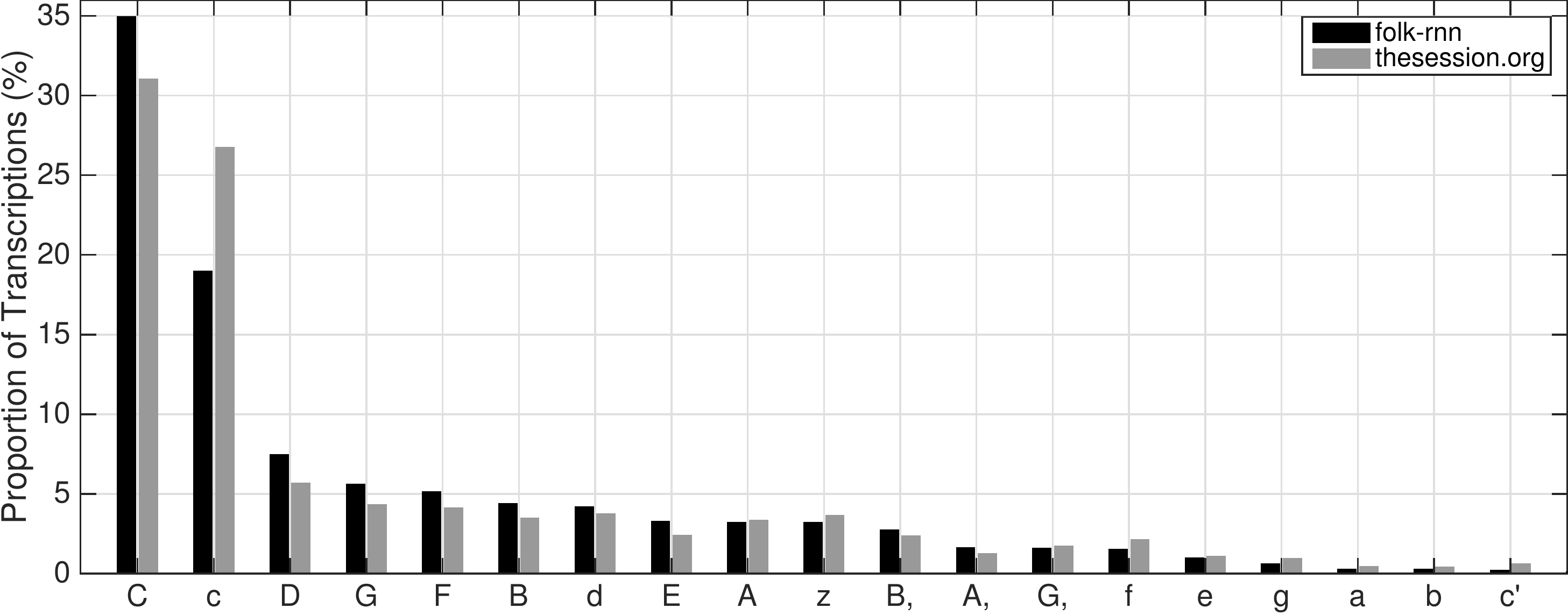}}
  \caption{Top: Distribution of the number of tokens in a transcription
  for the 6,101 transcriptions created by our {\em folk-rnn} system,
  compared with those in its (transposed) training dataset.
  Bottom: Proportion of transcriptions that conclude on a given pitch.}
  \label{fig:folkrnnproportions}
  \vspace{-0.2in}
\end{figure}

\vspace{-0.1in}
\subsection{Statistical analysis of outputs}
Comparing the descriptive statistics of system output
with those of its training data is a straightforward 
way of assessing its internal model,
but its relevance to the experience of music is highly questionable.
We take our {\em folk-rnn} system and have it generate 6,101 full transcriptions.
The proportions of meters and modes are 
close to those in the training dataset.
Figure \ref{fig:folkrnnproportions} shows the proportion of 
transcriptions of a particular token lengths,
and the proportion ending with a particular pitch.
The end pitch distributions appear to match between the two,
but not transcription token length.
We do not currently know the reason for this.
We also find (by looking at the occurrence of repeat signs)
that about 68\% of the {\em folk-rnn} transcriptions use measure tokens 
creating a structure AABB with each section being 8 bars long;
54\% of the transcriptions in 
the training data have this structure.
This kind of structure is common in Irish folk music \cite{Hillhouse2005a}.
When it comes to errors,
16 generated transcriptions have the token {\tt |1} (first ending) followed by {\tt |1} instead of {\tt |2};
and 6 have just {\tt |1} or {\tt |2} specified.
Three transcriptions have incompletely specified chords, 
i.e., {\tt ]} appears without an accompanying {\tt [}.
(We corrected such problems 
when creating the training data for this model.)

%
\vspace{-0.1in}
\subsection{Musical analysis of outputs}
\begin{figure}[t]
\centering
\includegraphics[width=1\columnwidth]{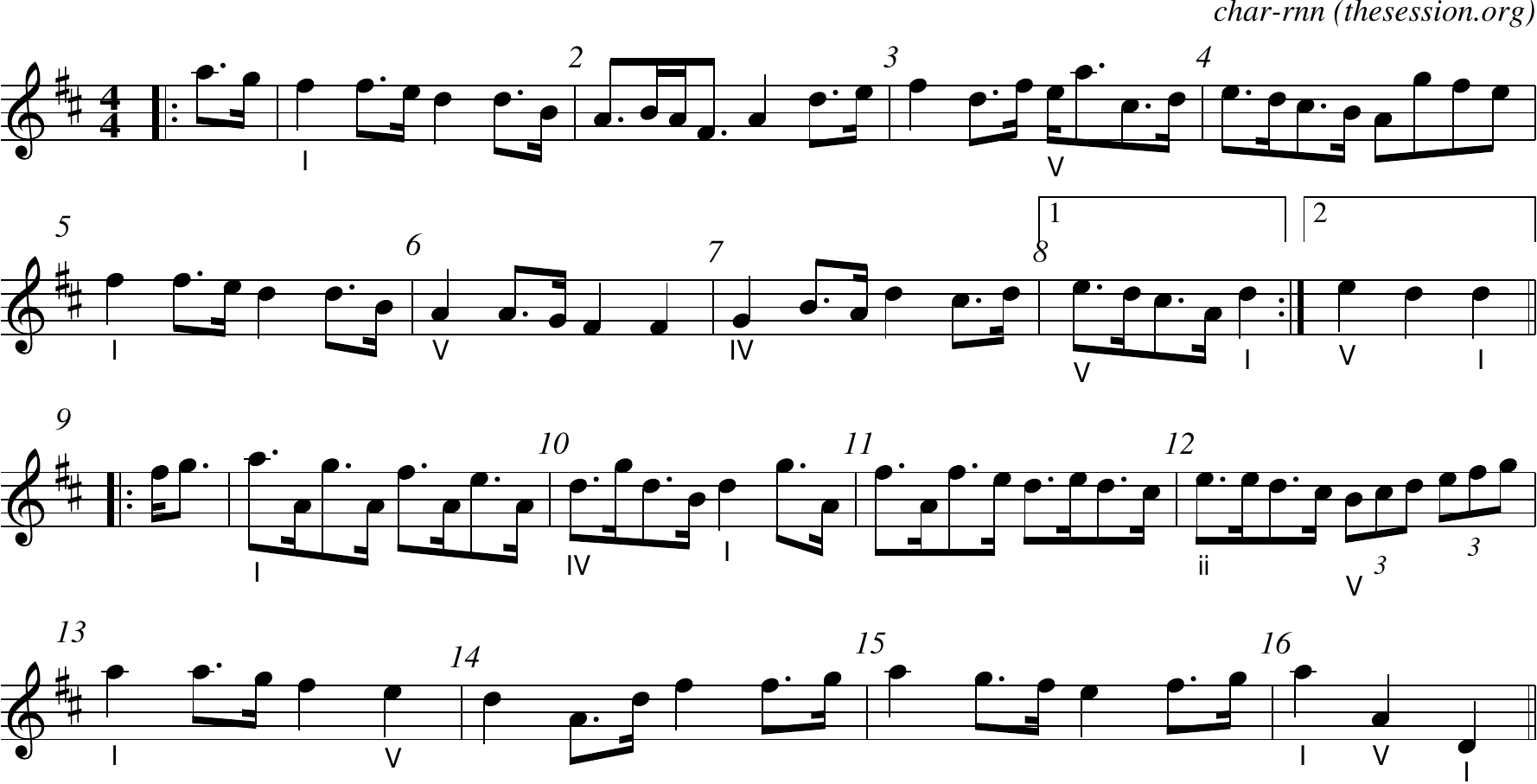}
\vspace{-0.2in}
\caption{Notation of ``The Mal's Copporim,'' to which we add implied harmonies.}
\label{fig:malscopporim}
\vspace{-0.2in}
\end{figure}

We generated 72,376 tune transcriptions from our {\em char-rnn} model, 
and automatically synthesised 35,809 of them (stopping only because of space limitations).\footnote{We 
use {\tt abc2midi} to convert each transcription to midi, 
and then process the midi using {\tt python-midi} to humanise it
for each of several randomly selected instruments, e.g., fiddle, box,
guitar/banjo, and drums, and then use {\tt timidity}, {\tt sox} and {\tt lame}
to synthesise, master, and compress as mp3.}
We used these results to create ``The Endless Traditional Music Session,''\footnote{\url{http://www.eecs.qmul.ac.uk/~sturm/research/RNNIrishTrad/index.html}}
which cycles through the collection in sets of seven 
randomly selected transcriptions every five minutes.
We shared this with the online community of {\tt thesession.org}.
One user listened to several, and identified the example below, saying,
``In the tune below, the first two phrases are quite fun as a generative idea to `human-compose' the rest of it! 
I know that's not quite the point of course. 
Still had fun trying the opening of this one on the harp.''
Here is the exact output of our {\em char-rnn} model
(notated in Fig. \ref{fig:malscopporim} with implied harmonies):\footnote{The system 
has in fact learned to create a title field for each transcription it produces 
because we include it in the training data for our {\em char-rnn} model.}

\vspace{-0.05in}
{\small
\begin{verbatim}
T: Mal's Copporim, The
M: 4/4
L: 1/8
K: Dmaj
|: a>g | f2 f>e d2 d>B | A>BA<F A2 d>e | f2 d>f e<ac>d | e>dc>B Agfe |
f2 f>e d2 d>B | A2 A>G F2 F2 | G2 B>A d2 c>d |[1 e>dc>A d2 :|[2 e2 d2 d2 ||
|: f<g | a>Ag>A f>Ae>A | d>gd>B d2 g>A | f>Af>e d>ed>c | e>ed>c (3Bcd (3efg |
a2 a>g f2 e2 | d2 A>d f2 f>g | a2 g>f e2 f>g | a2 A2 D2 ||
\end{verbatim}}\vspace{-0.05in}

\normalsize
\renewcommand{\abcwidth}{\dimexpr1\linewidth\relax}


Looking at this output
as a composition teacher would the work of a student, 
we find no glaring mistakes: 
all measures have correct durations
with accounting for the two pickup bars.
Only the repeat sign at the beginning of the turn is unbalanced. 
We see that the piece is firmly in D major (but see discussion of harmony below), 
and each section ends with a resolution, 
the most strong being the conclusion. 
The melody appropriately slows down at these end points. 
The piece shows a structure very common to traditional Irish music \cite{Hillhouse2005a}: 
a repeated 8 bar ``tune'' followed by a repeated 8 bar ``turn.''  
This is one point at which to suggest a change:
just as for the tune, give the turn two endings, 
making the one already there the last, 
and compose a less conclusive resolution as the first.

Looking at the melodic characteristics in both the tune and turn, 
the dominant contour is the descent.
The 3 stepwise notes beginning the piece, along with their rhythm, 
form a basic idea that is repeated verbatim or in transposition in several places. 
The piece shows clear use of repetition and variation: 
the turn keeps the dotted rhythm of the tune, but with a new melodic idea (for the first part of the phrase).
The dotted rhythm is repeated often but also varied. 
The occasional iamb adds variety and keeps the melody from becoming too monotonous, 
without breaking the strong metric character,
but that idea is abandoned after the first 3 measures. 
While it serves well in m. 2\&3,
the iamb variety in the upbeat to the turn is less effective.

The tune and turn sound related, 
with the turn opening with a variation of the stepwise motion of the tune. 
Measures 9\&10 in the turn vary bars 3 and 4 of the tune; 
and m. 13 in the turn recalls the beginning of the tune and its basic idea.
Overall, the turn sounds rather aimless in its last half, 
and the giant leaps in the final bar are unexpected given the gradual motion 
in most of the piece. 
Here is a second point at which we can improve the composition: 
make bar 5 of the turn more closely related to its first bar,
and change the rhythm of its second bar to that of the tune. 
The giant leaps in the last bar should be better prepared 
by the new first ending of the first suggestion above. 
Finally, in m. 6, change trochee rhythm to iamb 
and drop the second F-sharp to the D.\footnote{For example, {\tt A>B A<G F2 D2}.}


The transcription may be monophonic, but harmony is implicit in the melody.
(Chordal accompaniment became prevalent in session music since the early part of the 20th century
\cite{Hillhouse2005a}.)
In this piece, I (Dmajor) is the most common,
(e.g., m. 1-3) with V (Amajor) appearing as well (e.g., m. 3\&4),
and IV (Gmajor) appearing in m. 10.
There are some awkward harmonic moments: 
the V seems to arrive half a bar too early in m. 3; 
the first half of m. 10 is IV, but does one switch to V for the last beat,
or keep IV and ignore the melodic A? 
The harmony in m. 12 could be ii (Eminor) --- the only minor chord in the piece --- 
which leaves m. 13 with a V-I cadence but to a weak beat. 
The second half of the turn is quite static harmonically,
which contributes to its aimless quality.
That is a third point where we can improve the composition.\footnote{
One possibility is to change m. 13\&14 to {\tt a2 a>g f>A e>A | d2 A>d e2 f>g}.}

%

One might ask, in its generation of ``The Mal's Copporim'',
whether the system is just reproducing portions of its training dataset.
One characteristic element is the scalar run in the last half of m. 12.
We find this appears 13 times in 9 training transcriptions,
and in only three is it followed by the high A.
Another characteristic pattern is m. 9,
which appears (transposed) in only one training transcription,\footnote{``Underwood'' 
\small\url{https://thesession.org/tunes/5677}}
but in the context of v (minor), and followed by a measure quite different
from that in ``The Mal's Copporim''.
Another characteristic element is the ending measure,
which is not present in the training transcriptions.
We find only one instance of m. 2,\footnote{Version 
3 of ``Durham Rangers'' \small\url{https://thesession.org/tunes/3376}} 
but no instances of m. 3\&4.

\vspace{-0.1in}
\subsection{Music composition with the generative systems}
We now describe an instance of using our {\em char-rnn} system
to assist in the composition of a new piece of music. 
The process begins by seeding the system with the transcription of an idea,
judging and selecting from its output, 
and seeding anew with an expanded transcription.
We initialise the model with the following seed,
which includes two bars: 

{\small
\begin{verbatim}
T: Bob's Idea
M: 4/4
L: 1/8
K: Cmaj
|: CcDB E^A=AF | d2 cB c2 E2 |
\end{verbatim}}\vspace{-0.05in}

\normalsize
\noindent It generates 1000 new characters, which include 18 measures following the seed
to finish the tune.
We notate a portion of this below with the seed (m. 1\&2):
\includegraphics[width=1\columnwidth]{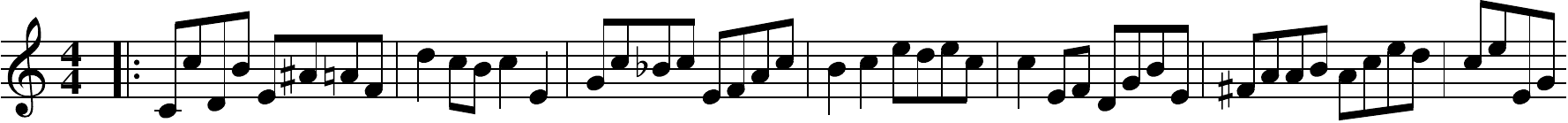}

\normalsize
\noindent
We keep the measure following the seed,
compose another measure that varies the m. 2,
and seed the system with those four measures.
The system then produces two four-measure endings: 

\noindent\includegraphics[width=1\columnwidth]{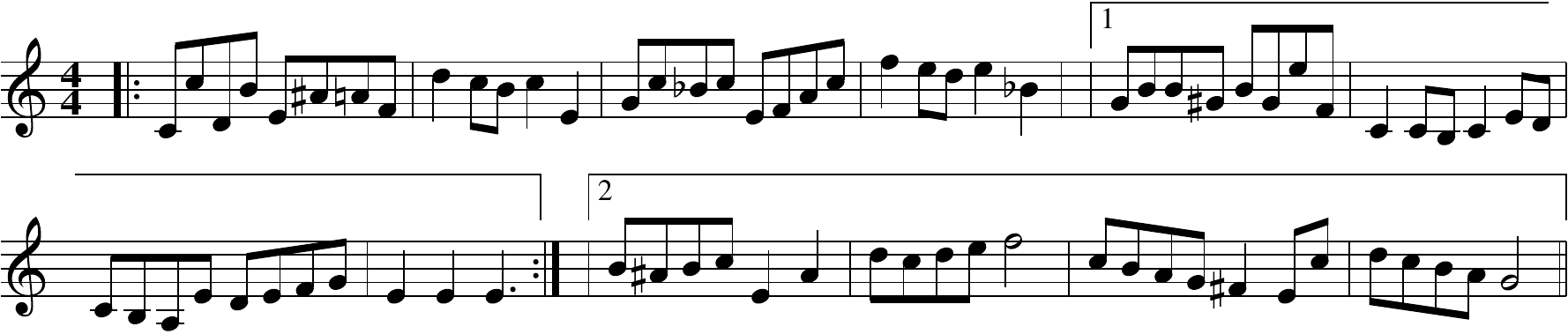}

\normalsize
\noindent We keep the music of the second ending,
and seed the system with

{\small
\begin{verbatim}
T: Bob's Idea
M: 4/4
L: 1/8
K: Cmaj
|: CcDB E^A=AF | d2 cB c2 E2 | Gc_Bc EFAc | f2 ed e2 _B2 |
B^ABc E2 A2 | dcde f4 | cBAG ^F2 Ec | dcBA G4 |
\end{verbatim}}

\normalsize
\noindent This produces 8 more measures, 
a few of which we notate below (m. 9-11):

\noindent\includegraphics[width=1\columnwidth]{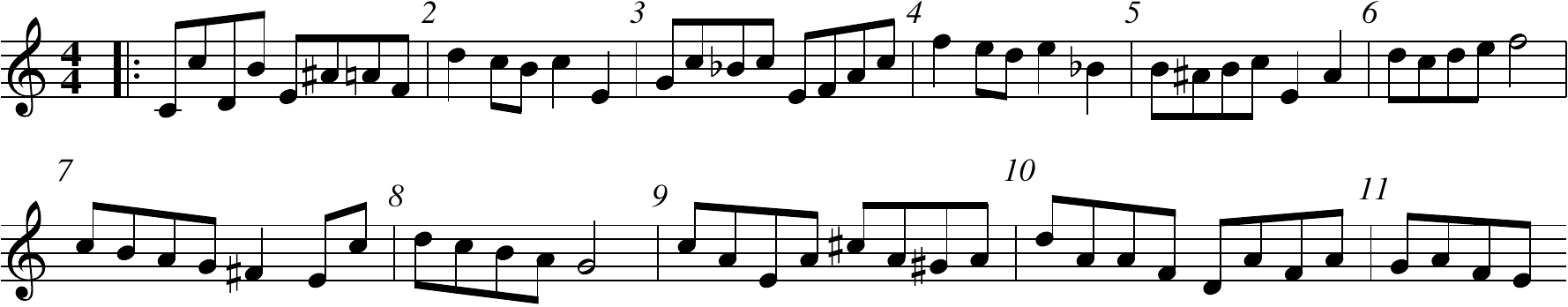}

\begin{figure}
\centering
\includegraphics[width=1\columnwidth]{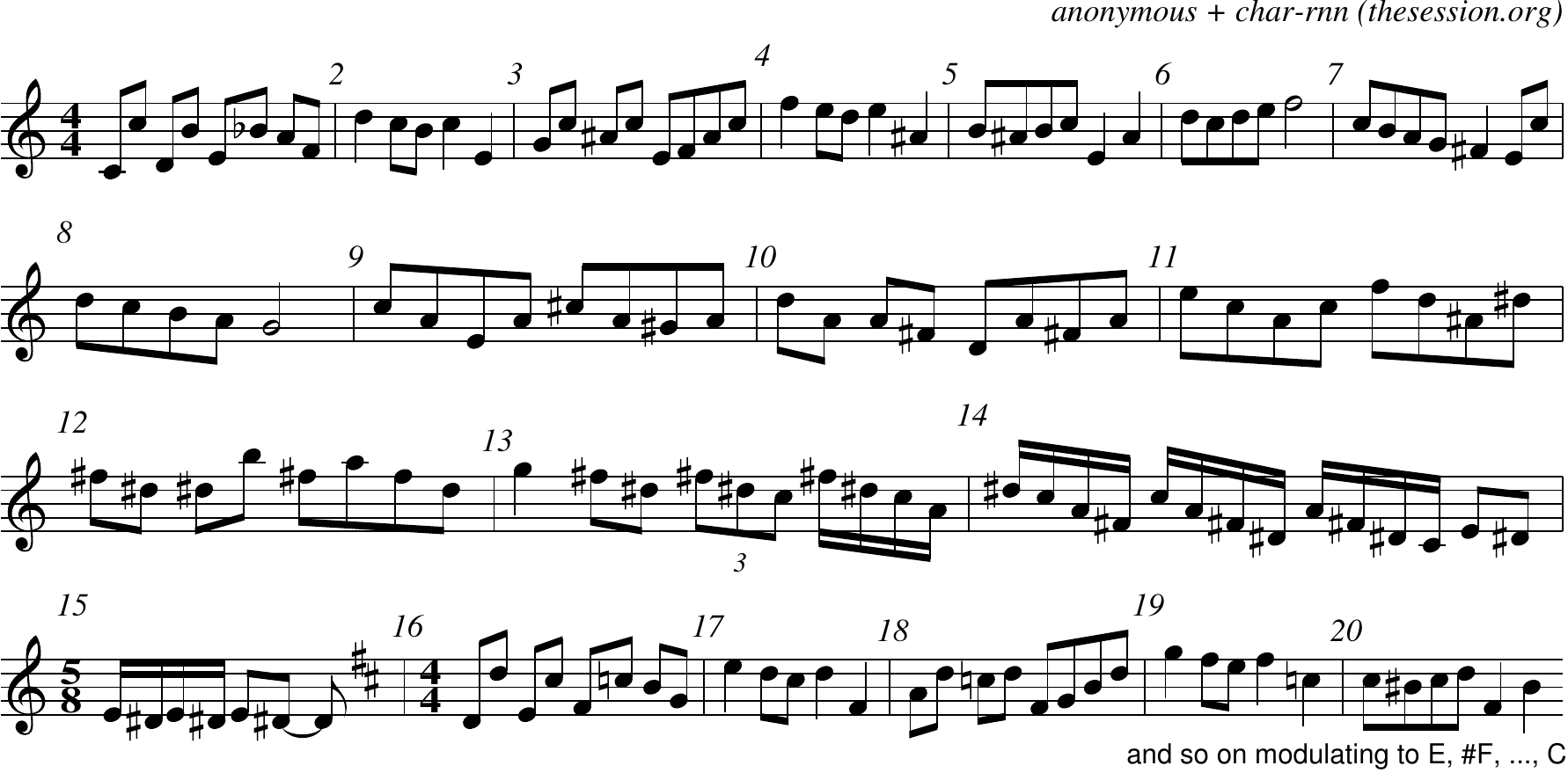}
\vspace{-0.2in}
\caption{The beginning of ``The March of Deep Learning'',
composed with assistance from the {\em char-rnn} model,
is quite different to the kind of music in the training data.}
\label{fig:marchofdeeplearning}
\vspace{-0.2in}
\end{figure}

\normalsize
\noindent We keep m. 9\&10, vary them to create two new bars,
then compose a few more measures to modulate to the V of V,
and then repeat the first 15 measures transposed a whole step up.
With a few more edits, we have composed ``The March of Deep Learning'',
Fig. \ref{fig:marchofdeeplearning},
which sounds quite different from the music in the 
training data transcriptions. 

\vspace{-0.1in}
\section{Discussion and reflection}\vspace{-0.1in}
The immediate practical aim of our work 
is to create music transcription models 
that facilitate music composition,
both within and outside particular conventions.
Toward this end, we have built two different kinds of generative systems
using deep learning methods and a large number of 
textual transcriptions of folk music,
and demonstrated their utility from three perspectives.
We compare the statistics of the generated output 
to those of the training material.
We analyse a particular transcription generated by 
one of the systems (notated in Fig. \ref{fig:malscopporim})
with respect to its merits and weaknesses as a composition,
and how it uses conventions found in traditional Celtic music.
We use one of the systems to help compose
a new piece of music (notated in Fig. \ref{fig:marchofdeeplearning}).\footnote{The
reason why we use {\em folk-rnn} for the first part and not the others
is purely because our preliminary experiments with LSTM networks
involved {\em char-rnn}. 
Our results led us to refine the transcription vocabulary 
and training regimen for {\em folk-rnn}.}

\begin{figure}
\centering
\includegraphics[width=1\columnwidth]{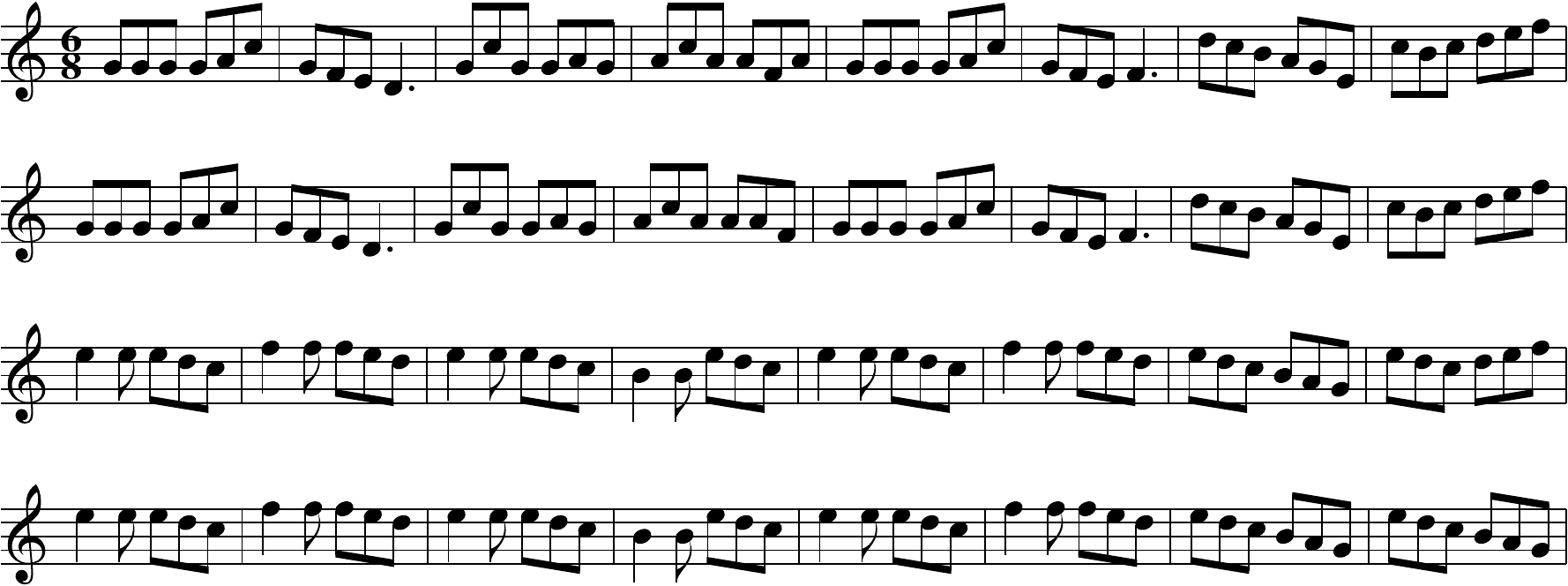}
\vspace{-0.2in}
\caption{Notated output of a {\em folk-rnn} model trained on transcriptions
with repetitions made explicit.}
\label{fig:worepetition}
\vspace{-0.2in}
\end{figure}

The statistics of the output of the {\em folk-rnn} system suggest 
that it has learned to count, in terms of the number of
notes per measure in the various meters present in the dataset.
This is consistent with previous findings about RNN \cite{Gers2000a}.
We can also see the distribution of pitches 
agree with that of the training data.
The {\em folk-rnn} system seems to have learned about 
ending transcriptions on the tonic;
and using measure tokens to create transcriptions with an AABB
structure with each section being 8 measures long.
In our latest experiments, 
we trained a {\em folk-rnn} system with transcriptions 
spelling out repeated measures 
(replacing each repeat sign with the repeated material). 
We find that many of the generated transcriptions (see Fig. \ref{fig:worepetition}) 
adhere closely to the AABB form, 
suggesting that this system is learning about repetition 
rather than where the repeat tokens occur.

A statistical perspective, however, is only able to reflect 
how well the learning algorithm has divined specific information 
about the training dataset to produce ``valid'' ABC output.
To learn more specific information about 
how well these systems can facilitate music composition,
we look at the level of individual transcriptions.
We take on the role of a composition teacher
assessing the work of a student.
While the question of creativity and composition teaching 
is not without contention (for example, \cite{Lupton2010} and \cite{Czernowin2012}), 
criteria such as creativity, imagination, originality and innovation 
are used in many music department when marking composition assignments.
Therefore, we can consider the perspective of a composition teacher 
a form of expert opinion on the ability of these systems,
but be careful to acknowledge two things: 
1) there is an inherited bias in Western musical culture
with regards to the importance of the personal voice;
2) while stylistic awareness informs the discussion of a music composition,
adherence to the conventions of a style 
is often not the primary focus of the ensuing discussion.



``The Mal's Copporim'' (notated in Fig. \ref{fig:malscopporim}),
is a very plausible music transcription that is nearly
``session-ready''. 
Through our own audition of many hundreds of results, 
we also find others that have similar plausibility. 
Certainly, our systems produce many transcriptions
that are much less plausible as well; 
and of course judging a transcription as plausible is, naturally, subjective;
but the argument we are making here is that these systems 
are producing music transcriptions that appear to 
be musically meaningful (repetition, variation, 
melodic contour, structure, progression, resolution).
We cannot dispense with the need for a curator 
to identify good and poor output;
or for a composer/performer to correct or improve an output.

The role of the composer is clear when we apply our system
to create a new piece of music in Sec. 4.3.
Our intention behind seeding the system
with the opening two bars of 
(Fig. \ref{fig:marchofdeeplearning}) 
is to see how the system 
responds to an input that does not adhere closely to 
the stylistic conventions in its training data.
Is it able to apply pattern variations even 
when the input pattern isn't very close to the learned material?
Through our experience, we find the knowledge embedded within the system
translates into this different context with the guiding hand of the composer.
Within this relatively restricted approach to composition,
we find our systems useful for assisting in  
music material generation that goes in directions we thought little to take.

Our work so far has merely
examined the ability of these deep learning methods
for modeling ABC transcriptions,
but further work is clear.
First, we will elicit discussions from {\tt thesession.org} community 
about the transcriptions produced by {\em folk-rnn},
and how they can be improved with respect to
stylistic and performance conventions.
Hillhouse \cite{Hillhouse2005a} mentions the openness of session musicians 
to incorporate new tunes into their performance repertoire,
and so we are interested to see if any incorporate some of our results.
Second, we will conduct interviews with session musicians 
to analyse {\em folk-rnn} transcriptions for their 
adherence to stylistic conventions, and how the experts would
change the transcriptions to better fit the style.
This will provide opportunities to improve the transcription model.
Third, we will build an interface such that 
users can explore the system for composing new music (much the way
we applied it in Sec. 4.3), and then measure 
how well it facilitates composition.
We also seek ways to adapt the models to other kinds of stylistic conventions,
and to analyse the significance of model parameters and network layers
to the musical knowledge implicit in the dataset.

\vspace{-0.1in}
\section{Conclusion}\vspace{-0.1in}
Facilitated both by the availability of data,
and the excellent reproducibility of research in deep learning,
our work extends past research in applying RNN and LSTM networks
to music modeling and composition 
\cite{Todd1989,Mozer1994a,Chen2001,Eck2002a,Franklin2006a,Eck2008a}
by virtue of size:
whereas past work has used up to only a few hidden layers
of a few dozen units, and a few hundreds of training examples,
to generate only few example sequences,
we have built networks containing thousands of units
trained on tens of thousands of training examples,
and generated tens of thousands of transcriptions.
We explore the learned models in several ways.
In addition to a comparison of the statistics of the generated transcriptions
and the training data,
we employ critical perspectives that are relevant to our aims:
to create music transcription models 
that facilitate music composition,
both within and outside particular conventions.

We make no claims that we are modelling music creativity \cite{Wiggins2009a}.
As they stand, these models are black boxes
containing an agent that uses 
probabilistic rules to arrange tokens \cite{Searle1980a}.
Curation, composition and performance are required 
to make the generated transcriptions become music. 
However, at the level of the transcriptions,
we find the collection of results to have a consistency in
plausibility and meaningful variation.
These LSTM networks are able to take a transcribed musical idea 
and transform it in meaningful ways.
Furthermore, our models seem quite applicable 
in the context of traditional Celtic music practice
because the creative practice of practitioners
lies in their ability to arrive at novel recombinations 
of familiar elements \cite{Cowdery1990a}.
Discovering a good balance between consistency and variation 
is part of the development of a composer's inner monitor 
and is a contributing factor to a composer's own style.
That presents a unique point at which our system could positively contribute.
However, it is still up to the composer to learn when and how to 
bend or break the rules to create music of lasting interest.
The application of machine learning is no substitute.

\vspace{-0.1in}
\bibliographystyle{plain}
\bibliography{../bibliographies/BibAnnon,../bibliographies/genre}

\begin{thebibliography}{10}

\bibitem{Bergstra2010}
J.~Bergstra, O.~Breuleux, F.~Bastien, P.~Lamblin, R.~Pascanu, G.~Desjardins,
  J.~Turian, D.~Warde-Farley, and Y.~Bengio.
\newblock Theano: A {CPU} and {GPU} math expression compiler.
\newblock In {\em Proc. Python for Scientific Computing Conf.}, June 2010.

\bibitem{Boulanger-Lewandowski2012a}
N.~Boulanger-Lewandowski, Y.~Bengio, and P.~Vincent.
\newblock Modeling temporal dependencies in high-dimensional sequences:
  Application to polyphonic music generation and transcription.
\newblock In {\em Proc. Int. Conf. Machine Learning}, 2012.

\bibitem{Boulanger-Lewandowski2014}
N.~Boulanger-Lewandowski, G.~J. Mysore, and M.~Hoffman.
\newblock Exploiting long-term temporal dependencies in {NMF} using recurrent
  neural networks with application to source separation.
\newblock In {\em Proc. Int. Conf. Acoustics, Speech, Signal Process.}, pages
  6969--6973, May 2014.

\bibitem{Chen2001}
C.~J. Chen and R.~Miikkulainen.
\newblock Creating melodies with evolving recurrent neural networks.
\newblock In {\em Proc. Int. Joint Conf. Neural Networks}, pages 2241--2246,
  2001.

\bibitem{Coca2011}
A.~E. Coca, R.~A.~F. Romero, and L.~Zhao.
\newblock Generation of composed musical structures through recurrent neural
  networks based on chaotic inspiration.
\newblock In {\em Int. Conf. Neural Networks}, pages 3220--3226, July 2011.

\bibitem{Cowdery1990a}
J.~R. Cowdery.
\newblock {\em The melodic tradition of Ireland}.
\newblock Kent State Uni. Press, 1990.

\bibitem{Czernowin2012}
C.~Czernowin.
\newblock Teaching that which is not yet there (stanford version).
\newblock {\em Contemporary Music Review}, 31(4):283--289, 2012.

\bibitem{Deng2014}
L.~Deng and D.~Yu.
\newblock {\em Deep Learning: Methods and Applications}.
\newblock Now Publishers, 2014.

\bibitem{Dolson1989a}
M.~Dolson.
\newblock Machine tongues {XII}: Neural networks.
\newblock {\em Computer Music J.}, 13(3):3--19, 1989.

\bibitem{Eck2008a}
D.~Eck and J.~Lapamle.
\newblock Learning musical structure directly from sequences of music.
\newblock Technical report, University of Montreal, 2008.

\bibitem{Eck2002a}
D.~Eck and J.~Schmidhuber.
\newblock Learning the long-term structure of the blues.
\newblock In {\em Proc. Int. Conf. on Artificial Neural Networks}, 2002.

\bibitem{Franklin2006a}
J.~A. Franklin.
\newblock Recurrent neural networks for music computation.
\newblock {\em J. Computing}, 18(3):321--338, 2006.

\bibitem{Gatys2015}
L.~A. Gatys, A.~S. Ecker, and M.~Bethge.
\newblock A neural algorithm of artistic style.
\newblock {\em CoRR}, abs/1508.06576, 2015.

\bibitem{Gers2000a}
F.~A. Gers and J.~Schmidhuber.
\newblock Recurrent nets that time and count.
\newblock In {\em Proc. Int. Joint Conf. on Neural Networks}, 2000.

\bibitem{Graves2013b}
A.~Graves.
\newblock Generating sequences with recurrent neural networks.
\newblock {\em CoRR}, abs/1308.0850, 2013.

\bibitem{Graves2013a}
A.~Graves, A.-R. Mohamed, and G.~Hinton.
\newblock Speech recognition with deep recurrent neural networks.
\newblock In {\em Proc. Int. Conf. Acoustics, Speech, Signal Process.}, pages
  6645--6649, 2013.

\bibitem{Griffith1999}
N.~Griffith and P.~M Todd.
\newblock {\em Musical networks: Parallel distributed perception and
  performance}.
\newblock MIT Press, 1999.

\bibitem{Hillhouse2005a}
A.~N. Hillhouse.
\newblock Tradition and innovation in {I}rish instrumental folk music.
\newblock Master's thesis, The University of British Columbia, 2005.

\bibitem{Hinton2012}
G.~Hinton, L.~Deng, D.~Yu, G.~E. Dahl, A.-R. Mohamed, N.~Jaitly, A.~Senior,
  V.~Vanhoucke, P.~Nguyen, and T.~N. Sainath.
\newblock Deep neural networks for acoustic modeling in speech recognition: The
  shared views of four research groups.
\newblock {\em Signal Process. Mag.}, 29(6):82--97, 2012.

\bibitem{Hochreiter1998}
S.~Hochreiter.
\newblock The vanishing gradient problem during learning recurrent neural nets
  and problem solutions.
\newblock {\em Int. J. Uncertain. Fuzziness Knowl.-Based Syst.}, 6(2):107--116,
  April 1998.

\bibitem{Hochreiter1997}
S.~Hochreiter and J.~Schmidhuber.
\newblock Long {Short}-{Term} {Memory}.
\newblock {\em Neural Computation}, 9(8):1735--1780, November 1997.
\newblock bibtex: Hochreiter1997.

\bibitem{Humphrey2013}
E.~Humphrey, J.~P. Bello, and Y.~LeCun.
\newblock Feature learning and deep architectures: New directions for music
  informatics.
\newblock {\em J. Intell. Info. Systems}, 41(3):461--481, 2013.

\bibitem{Kereliuk2015a}
C.~Kereliuk, B.~L. Sturm, and J.~Larsen.
\newblock Deep learning and music adversaries.
\newblock {\em IEEE Trans. Multimedia}, 17(11):2059--2071, Sep. 2015.

\bibitem{Krizhevsky2012}
A.~Krizhevsky, I.~Sutskever, and G.~E. Hinton.
\newblock Imagenet classification with deep convolutional neural networks.
\newblock In {\em Proc. NIPS}, pages 1097--1105, 2012.

\bibitem{LeCun2015}
Y.~{LeCun}, Y.~Bengio, and G.~Hinton.
\newblock Deep learning.
\newblock {\em Nature}, 521(7553):436--444, 2015.

\bibitem{Lee2009}
H.~Lee, Y.~Largman, P.~Pham, and A.~Y. Ng.
\newblock Unsupervised feature learning for audio classification using
  convolutional deep belief networks.
\newblock In {\em Proc. Neural Info. Process. Systems}, pages 1096--1104, 2009.

\bibitem{Leman1992}
M.~Leman.
\newblock Artificial neural networks in music research.
\newblock In Marsden and Pople, editors, {\em Computer Representations and
  Models in Music}. Academic Press, 1992.

\bibitem{Lupton2010}
M.~Lupton and C.~Bruce.
\newblock Craft, process and art: Teaching and learning music composition in
  higher education.
\newblock {\em British J. Music Education}, 27(3):271--287.

\bibitem{Mozer1994a}
M.~C. Mozer.
\newblock Neural network composition by prediction: Exploring the benefits of
  psychophysical constraints and multiscale processing.
\newblock {\em Cog. Science}, 6(2\&3):247--280, 1994.

\bibitem{Pascanu2013a}
R.~Pascanu, T.~Mikolov, and Y.~Bengio.
\newblock On the difficulty of training recurrent neural networks.
\newblock {\em J. Machine Learning Res.}, 28(3):1310--1318, 2013.

\bibitem{Searle1980a}
J.~Searle.
\newblock Minds, brains and programs.
\newblock {\em Behavioral \& Brain Sci.}, 3(3):417--57, 1980.

\bibitem{Sigtia2014}
S.~Sigtia and S.~Dixon.
\newblock Improved music feature learning with deep neural networks.
\newblock In {\em Proc. Int. Conf. Acoustics, Speech Signal Process.}, pages
  6959--6963, May 2014.

\bibitem{Spiliopoulou2011}
A.~Spiliopoulou and A.~Storkey.
\newblock Comparing probabilistic models for melodic sequences.
\newblock In {\em Proc. Machine Learn. Knowledge Disc. Data.}, pages 289--304,
  2011.

\bibitem{Sturm2014}
B.~L. Sturm, C.~Kereliuk, and A.~Pikrakis.
\newblock A closer look at deep learning neural networks with low-level
  spectral periodicity features.
\newblock In {\em Proc. Int. Workshop on Cognitive Info. Process.}, pages 1--6,
  2014.

\bibitem{Sutskever2014}
I.~Sutskever, O.~Vinyals, and Q.~V. Le.
\newblock Sequence to sequence learning with neural networks.
\newblock In {\em Proc. Neural Information Process. Systems}, pages 3104--3112,
  2014.

\bibitem{Todd1989}
P.~M. Todd.
\newblock A connectionist approach to algorithmic composition.
\newblock {\em Computer Music J.}, 13(4):27--43, 1989.

\bibitem{Todd1991a}
P.~M. Todd and G.~D. Loy.
\newblock {\em Music and connectionism}.
\newblock The MIT Press, 1991.

\bibitem{Tudor1992a}
D.~Tudor.
\newblock Neural network plus.
\newblock {\em (music score)}, 1992.

\bibitem{Wiggins2009a}
G.~A. Wiggins, M.~T. Pearce, and D.~M\"ullensiefen.
\newblock Computational modelling of music cognition and musical creativity.
\newblock In R.~T. Dean, editor, {\em The Oxford Handbook of Computer Music}.
  Oxford University Press, 2009.

\bibitem{Yang2011b}
X.~Yang, Q.~Chen, S.~Zhou, and X.~Wang.
\newblock Deep belief networks for automatic music genre classification.
\newblock In {\em Proc. INTERSPEECH}, pages 2433--2436, 2011.

\bibitem{Zhang2014}
C.~Zhang, G.~Evangelopoulos, S.~Voinea, L.~Rosasco, and T.~Poggio.
\newblock A deep representation for invariance and music classification.
\newblock In {\em Proc. Int. Conf. Acoustics, Speech Signal Process.}, pages
  6984--6988, May 2014.

\end{thebibliography}

\end{document}